\documentclass[journal=jacsat,manuscript=article]{achemso}

\usepackage{amsmath}
\usepackage{amssymb}
\usepackage{amsfonts}
\usepackage{color}
\usepackage{bm}
\usepackage[english]{babel}
\usepackage{hyperref}
\usepackage{siunitx}
\usepackage{graphicx}
\usepackage{enumerate}
\usepackage[version=3]{mhchem} 
\usepackage{physics}
\usepackage{setspace}
\usepackage{geometry}
\usepackage{polski}
\renewcommand{\vec}[1]{\ensuremath{\mathbf{#1}}}

\newcommand{\widefigurewidth}{0.45\textwidth}
\newcommand{\kB}{\ensuremath{k_{\rm B}}}
\newcommand{\kb}{\ensuremath{\kB}}
\newcommand{\kT}{\ensuremath{\kb T}}

\author{Jiale Shi}
\affiliation{%
  Department of Chemical and Biomolecular Engineering, %
  University of Notre Dame, %
  Notre Dame, IN 46556, USA
}

\author{Michael J. Quevillon}%
\affiliation{%
  Department of Chemical and Biomolecular Engineering, %
  University of Notre Dame, %
  Notre Dame, IN 46556, USA
}

\author{Pedro H. Amorim Valen\c{c}a}%
\affiliation{%
  Department of Chemical and Biomolecular Engineering, %
  University of Notre Dame, %
  Notre Dame, IN 46556, USA
}

\author{Jonathan K. Whitmer}
\email{jwhitme1@nd.edu}
\affiliation{%
  Department of Chemical and Biomolecular Engineering, %
  University of Notre Dame, %
  Notre Dame, IN 46556, USA
}
\altaffiliation{%
Department of Chemistry and Biochemistry, University of Notre Dame, Notre Dame, IN 46556, USA
}%

\title[]{Predicting Adhesive Free Energies of Polymer--Surface Interactions with Machine Learning}

\begin{document}

\begin{abstract}
Polymer-surface interactions are crucial to many biological processes and industrial applications. Here we propose a machine-learning method to connect a model polymer's sequence with its adhesion to decorated surfaces. We simulate the adhesive free energies of $20,000$ unique coarse-grained 1D sequential polymers interacting with functionalized surfaces and build support vector regression (SVR) models that demonstrate inexpensive and reliable prediction of the adhesive free energy as a function of the sequence. Our work highlights the promising integration of coarse-grained simulation with data-driven machine learning methods for the design of new functional polymers and represents an important step toward linking polymer compositions with polymer-surface interactions.
\end{abstract}

\maketitle

\section*{Introduction}
Polymer-surface interactions are critical to many industrial applications and biological processes.~\cite{kim2003epitaxial, chakraborty2001polymer, chakraborty2001disordered, xiu2020Inhibitors} The daily activity of writing on paper with inks composed of macromolecules is a ubiquitous case of polymer-surface interactions.~\cite{chakraborty2001polymer} Many industrial fabrication processes, such as using polymers to coat magnetic storage media, chips, and silicon capacitors, also involve polymer-surface interactions. As one example, Kim et al.~\cite{kim2003epitaxial} utilize the interactions of block copolymers with chemical patterning of surfaces to induce epitaxial self-assembly of block polymer domains, allowing for molecular-level control in top-down fabrication techniques. Additionally, many biological processes also involve what in essence are polymer-surface interactions. The interactions between heterochromatin and nuclear lamina affect the cell nuclear reorganization, reflecting cellular senescence processes.~\cite{chiang2019polymer} Vital biological processes such as intracellular signaling and incorporation of viruses into host cells,~\cite{xiu2020Inhibitors,wong2021sars} are initiated by a protein searching for and recognizing a specific receptor on a cell surface.  A small change to the sequence of the polymers affects their interactions and adhesive properties. For instance, mutations in the virus, like D164G, N501Y, and 501.V2 of COVID-19,~\cite{callaway2020making, Plante2021, hie2021learning}  change the spike protein structures and functionalities enabling the new spike protein to bind more easily with the ACE2 receptor, leading to a larger likelihood of infections. On a practical level, an understanding of the quantitative effects polymer sequences can have on surface adhesion is an essential ingredient for the design and synthesis of adhesive materials for tissues,~\cite{Annabi2014} where the surfaces to be attached may have significant compositional heterogeneity.

Several early theoretical and computational studies have examined the effects of polymer composition on polymer-surface interactions and molecular recognition.~\cite{ozboyaci2016modeling, chakraborty1998simple, muthukumar1995pattern, Muthukumar1998pattern} Chakraborty  and Bratko~\cite{chakraborty1998simple} utilized Monte Carlo simulation to study the adsorption of random heteropolymers (RHPs) on disordered multifunctional surfaces, finding that a sharp adsorption transition occurs when statistical pattern matching exists between the RHP sequence and the surface site distribution. Muthukumar~\cite{muthukumar1995pattern, Muthukumar1998pattern} performed studies utilizing both theoretical analysis and Monte Carlo simulations to study the interactions of a polyelectrolyte chain with a patterned surface of opposite charge, illustrating that the self-assembly of polymer molecules at patterned surfaces is largely affected by the charge density, the size of the pattern, and the Debye length. It is known that polymer structural properties, such as the radius of gyration, can also influence polymer-surface interactions.~\cite{Chauhan2021crowding} A coarse-grained statistical mechanical model of AB co-polymers interacting with stripes of A and B beads on a surface was performed by Kriksin et al.,\cite{kriksin2005adsorption} who found that the adsorption behavior strongly depends on the copolymer sequence distribution and the arrangement of selectively adsorbing regions on the substrate.~\cite{kriksin2005adsorption} While some of these studies emphasize the importance of polymer sequence in surface adhesion,~\cite{chakraborty1998simple} both qualitative and quantitative understanding of sequence design principles is lacking.~\cite{chakraborty1998simple} The formidable challenge here is that databases for polymer sequences and structures comparable to extant databases for comparable properties~\cite{Webb2020Target}, do not exist and are expensive to create.

Machine learning (ML) and artificial intelligence (AI) have emerged as powerful tools for physical science and engineering,~\cite{artrith2021best, dePablo2019, gormley2021machine, huang2020deepPurpose} highlighted by recent projects AlphaFold2~\cite{Jumper2021, Tunyasuvunakool2021} and RoseTTAFold.~\cite{Baek2021} Naturally this has opened the door to many investigations exploring the predictions of polymer structure and therefore function from the sequence information, as well as ``inverse-design'' research.~\cite{Webb2020Target, ma2020transfer, ma2018determining, perry2020macro, ferguson2021macro} For instance, Statt et al.\cite{statt2020model,statt2021unsupervised,reinhart2021opportunities} have investigated the sequence-dependent aggregation  behavior of sequence-defined macromolecules via the unsupervised learning method. Another interesting case is that Meenakshisundaram and co-workers ~\cite{Meenakshisundaram2017design} have designed sequence-specific copolymer compatibilizers using a genetic algorithm applied to a coarse-grained molecular dynamics model. In one important related example, Webb et al.~\cite{Webb2020Target} utilized a deep neural network (DNN)  to predict structural properties of sequence-controlled coarse-grained polymers just from the sequence information. These successful cases~\cite{statt2020model, statt2021unsupervised, Meenakshisundaram2017design, Webb2020Target} inspire us to utilize ML and AI to investigate the quantitative relationships between the adhesive free energies and the polymer sequence information.

In this work, we utilize biased molecular dynamics simulations to generate a database of free energies which connect the sequence polymer and composition of a patterned surface to their adhesive properties. From this database, we build an inexpensive surrogate model using support vector regression (SVR) which demonstrates reliable prediction of the adhesive free energy of the polymer-surface interaction as a function of provided polymer sequence information. Subsequently, we apply this model to design targeted sequences using a genetic algorithm. Finally, we illustrate how the polymer sequence can be manipulated to affect the adhesive free energy with the surface and how to do inverse-engineering based on the sequential model-based genetic algorithm.

\begin{figure}
	\begin{center}
	\includegraphics[width=\widefigurewidth]{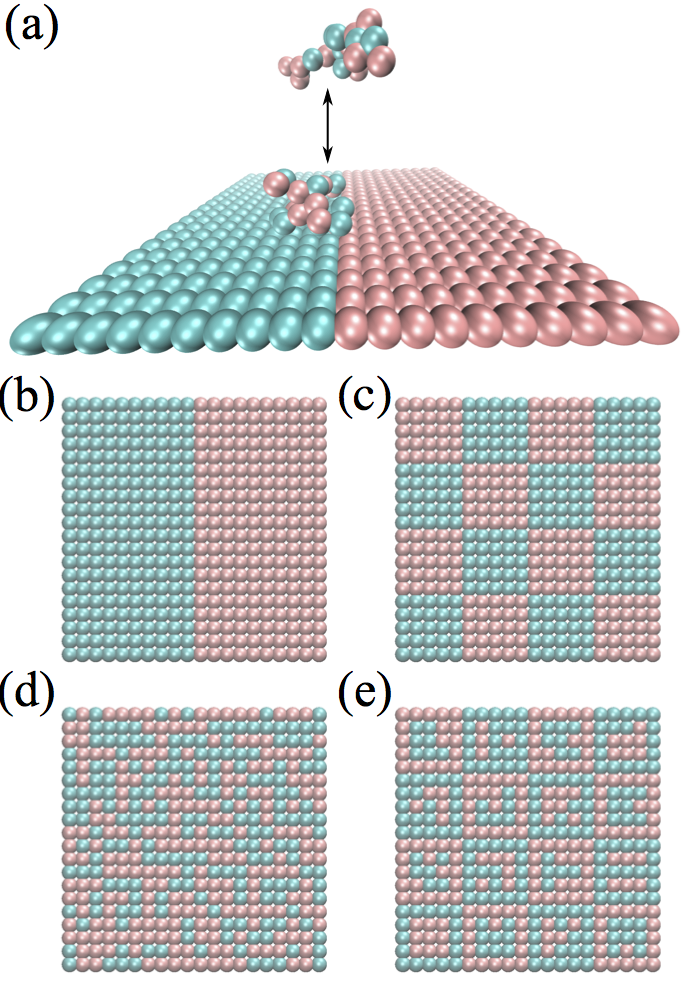}
	\end{center}
	\caption{(a) Schematic representation of our simulations involving polymer chain of defined sequence interacting with a patterned surface with a cubic simulation box whose side length $a=20\sigma$. The polymer chain and surface are both composed of two types of beads, A (red) and B (green). The polymer is modeled as a flexible 20-bead linear chain. The surface is holonomically constrained, with dimensions $20 \sigma \times 20 \sigma$ arranged in a simple square lattice with 400 beads. The four surfaces we examine (b-d) have different patterns with composition divided approximately equally between A and B beads. 
	(b) PS1, which is composed of half A beads and half B beads in two stripes. 
	(c) PS2, which is composed of 16 alternate small size squares ($5\sigma \times 5\sigma$) of A and B beads.
	(d) PS3, whose whole pattern is randomly generated from an assumption of equal probability for A and B beads . 
	(e) PS4, which is built upon PS2, but randomized within the interior of the $5\sigma \times 5\sigma$ squares.
	}
	\label{fig:model}
\end{figure}

\section*{Methods}

\subsection*{MD simulation}

We utilize model polymer chains containing 20 backbone beads based on the classic model of Kremer and Grest,~\cite{KremerGrestmodel1990} which has been widely utilize to investigate polymer interfacial properties.~\cite{Meenakshisundaram2017design,estridge2015diblock,lang2014combined,zhan2015coarse} The pair interaction between  beads is described via a 12--6 Lennard-Jones (LJ) potential:
\begin{equation}
    E^{ij}_{\rm LJ} = 4 \epsilon_{ij} \left[\left(\frac{\sigma}{r}\right)^{12} - \left(\frac{\sigma}{r}\right)^{6} \right]\;.
\end{equation}
where $\epsilon_{ij}$ sets the interaction energy between two type beads (red beads are A and green beads are B, $\epsilon_{\rm AA} = \epsilon_{\rm BB} = 1$, $\epsilon_{\rm AB} = 0.3 $ ), $\sigma$ sets the range of the interaction, and $r$ is the distance between two beads in dimensionless LJ units. The AA and BB Lennard-Jones interactions are truncated at a distance of $2.5 \sigma$, while the the AB Lennard-Jones interaction is truncated at a distance of $2^{1/6}\sigma$ so that it is purely repulsive. While simple, the construction of the model imposes an asymmetry in adhesion that can be optimized by searching over the sequence space of the polymers. Finally, bonds are handled via the finitely-extensive nonlinear elastic (FENE) potential\cite{GrestKremer1986finebond, stevens1993structure, kremer1990molecular, zhou2004cost} 

\begin{equation}
\begin{aligned}
    E_{\rm bond}  =& - \frac{1}{2} KR^2_{0} \ln \left[1 - \left(\frac{r}{R_{0}}\right)^2 \right] \;.
\end{aligned}
\end{equation}
\noindent where $K = 30 \epsilon/\sigma^2$ is the spring constant, $R_{0} = 1.5 \sigma$ is the maximum extent.

The patterned surface (PS) is $20 \sigma \times 20 \sigma$, containing $400$  beads type A or B, which have the same 12--6 LJ potential setting as the beads of the sequential polymer.  The PS's position is fixed at the plane $z=0$, and the distance between each bead is $\sigma$, shown in Fig.~\ref{fig:model}(a). As shown in Fig.~\ref{fig:model}(b-e), we investigate four different PSs to validate that our method is robust for surfaces with different patterns. 

\subsection*{Enhanced Sampling}
\begin{figure}
	\begin{center}
	\includegraphics[width=\widefigurewidth]{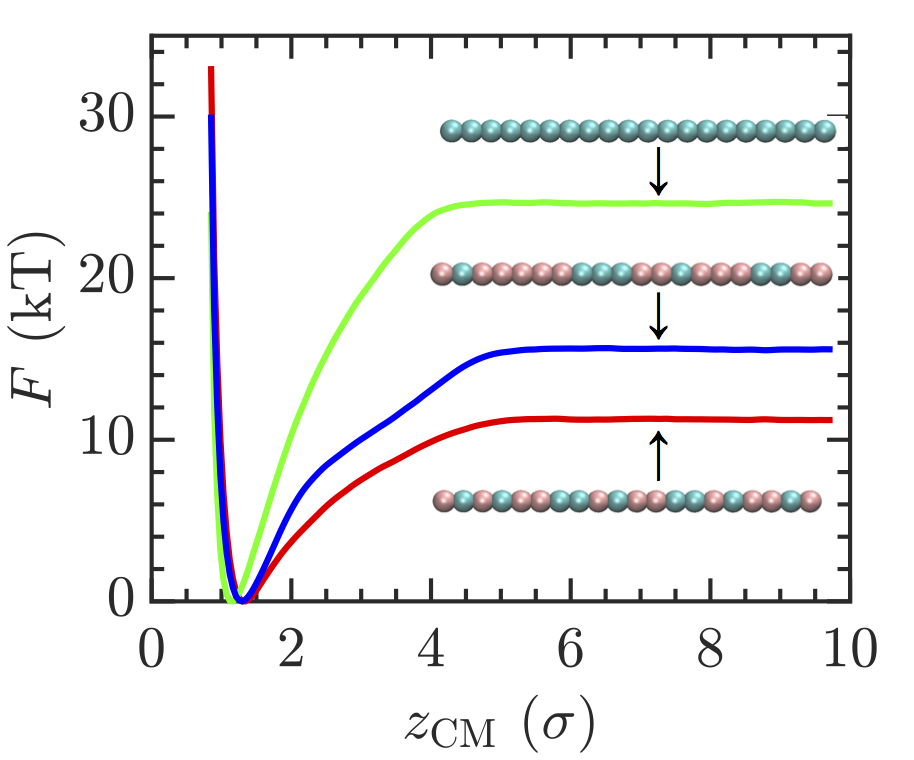}
	\end{center}
	\caption{Example potentials of mean force (PMFs) $F\ ({\rm in} \ \kT)$  plotted as a function of $z_{\rm  CM} \ ( {\rm in} \ \sigma)$ [$0.8\sigma,9.8\sigma$] for polymers with different sequences interacting with PS1. Generally, as  $z_{\rm CM}$ increases from $0.8\sigma$,  $F$ first decreases to a minimum value $F_{\rm min}$ because of the repulsive force between the polymer and the surface; then $F$ increases because of the attractive force between the polymer and the surface; finally when $z_{\rm CM} > 7$ approximately, $F$ becomes flat at the value of $F_{\rm o}$ as there is no interaction between the polymer and the surface. The adhesive energy for each sequential polymer chain interacting with the surface is $\Delta F  = F_{\rm o} - F_{\rm min}$. The influence of sequence on these quantities is illustrated by the insets, which show the sequence of polymer interacting with PS1 to obtain the given curve.
	}
	\label{fig:FES}
\end{figure}

Elucidating the adhesive free energies of sequential polymers with patterned surfaces requires efficient sampling of the rare events comprising removal and readhesion of a polymer to the interface, since significant energy barriers must be scaled to enable these rearrangements. Enhanced sampling calculations proceed by applying a bias to collective variables to speed up the exploration of the simulation systems. Collective variables (CVs), closely related to the concept of reaction coordinates, are a low-dimensional projection of the high-dimensional space of MD simulations, which can clearly distinguish reactants from products and quantify dynamical progress along the pathway from reactants to products.~\cite{peters2017reaction} Generally, this defines a vector valued function from the space of nuclear positions to the reduced CV space, $\vec{\xi}: \mathbb{R}^{3N} \rightarrow \mathbb{R}^{n}$, where $N$ is the number of atoms and $n$ the desired reduced dimensionality. For studying the adsorption process on a surface, it is typically sufficient to define a single collective variable. Here, we opt for a single CV ($d$),  the distance between the polymer chain's center of mass ($z_{\rm CM}$) and the surface ($z_{\rm surface} = 0$), 
\begin{equation}
    d = z_{\rm CM} - z_{\rm surface} = z_{\rm CM}\;.
\end{equation}
Here $d \equiv z_{\rm CM}$ since the patterned surface is located at $z=0$; we thus use $z_{\rm CM}$ as the CV for later descriptions. We obtain potentials of mean force (PMFs) for this coordinate using the multi-walker adaptive bias force (ABF) algorithm~\cite{Darve2008} as implemented in SSAGES.~\cite{sidky2018ssages} We choose to sample $z_{\rm CM}$ in range of $[0.8\sigma,\textbf{}9.8\sigma]$ with 90 bins. We use 4 walkers starting from different initial configurations, running each for a total of  $2\times10^7$ molecular dynamics timesteps to obtain a converged result. Example PMFs ($F$ vs. $z_{\rm CM}$) are plotted in Fig.~\ref{fig:FES}. Generally, as  $z_{\rm CM}$ increases from $0.8\sigma$, $F$ first decreases to a minimum value $F_{\rm min}$ because of the repulsive force between the polymer and the surface. Subsequently, then $F$ increases because of the attractive force between the polymer and the surface. finally, when $z_{\rm CM} > 7$ approximately, $F$ becomes flat at the plateau free energy in the noninteracting state $F_{\rm o}$ as there is no interaction between the polymer and the surface. We define the adhesive energy $\Delta F$ for each sequential polymer chain interacting with the PS as the difference between the plateaued noninteracting state $F_{\rm o}$ and the minimum state $F_{\rm min}$, $\Delta F  = F_{\rm o}- F_{\rm min}$,\cite{Chauhan2021crowding} and use this quantity to train the machine learning models explored in this paper. We note from Fig.~\ref{fig:FES} that PMF landscapes have similar shapes, but vary in magnitude for different polymer sequences. While each polymer--surface interaction can potentially have more subtle features, though $\Delta F$ captures the essential adhesive property.

\subsection*{Machine Learning}
We use support vector regression (SVR)~\cite{Drucker1997, bishop2006PRML, murphy2012machine}  to build a  model predicting polymer-surface interactions $\Delta F$ from limited polymer sequence information. The basic idea of a support vector machine (SVM)~\cite{cortes1995support, chang2011libsvm, bishop2006PRML, murphy2012machine} is to first map the data into a high dimensional space and then construct an optimal separating hyperplane in this space. The SVM thus constructed is then used to perform SVR. We utilize the SVR implementation in the open-source python package Scikit-learn~\cite{scikit-learn} using radial basis functions are for the regression. The settings of the optimized values of the regularization parameter ($C$) of the error term, the maximum error ($\epsilon$) that specifies the penalty-free area, and the kernel coefficient ($\gamma$) are stored in the Github repository described in section Code Availability. All the parameters mentioned above are optimized using 5-fold cross-validation.

\begin{figure}
	\begin{center}
	\includegraphics[width=\widefigurewidth]{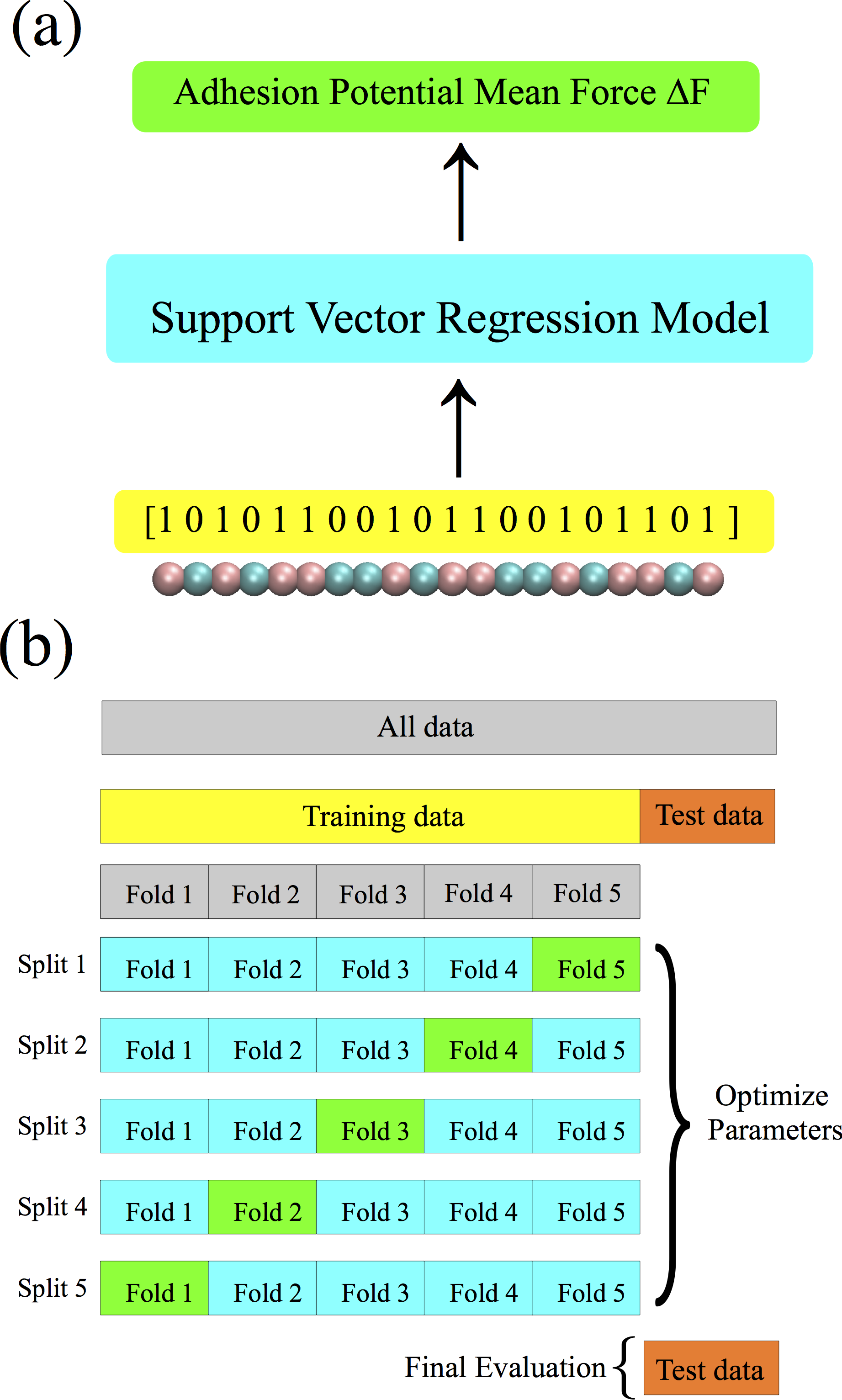}
	\end{center}
	\caption{(a) Schematic of our machine learning framework for predicting the polymer-surface interactions from sequence information. We use one-hot encoding to transfer the 20-bead-length 1D polymer sequence into a 20 dimensional vector using  where type A beads are 1 and type B beads are 0. The input is a 20 dimensional vector, while the output is the corresponding adhesive free energy $\Delta F$. The support vector regression ML models used in this work are dependent on the surface patterns. We investigate four different patterned surfaces in this work and train each corresponding individual support vector regression machine learning model. (b) We separate the $20,000$ unique polymer sequence into train data ($80\%$, $16,000$) and test data ($20\%$, $4,000$). Inside the training data, to avoid overfitting, we use 5-fold cross-validation to optimize the SVR model's parameters (the regularization parameter $C$, the maximum error $\epsilon$ and the kernel coefficient $\gamma$).~\cite{cortes1995support, chang2011libsvm, bishop2006PRML} Next, we train on the whole training data with the optimized SVR model. Finally, we evaluate the model's performance on the remaining test data which has not been used in the 5-fold cross-validation.
	}
	\label{fig:ML_framework}
\end{figure}

There are different types of polymer representations, like one-hot encoding,~\cite{Webb2020Target,ma2018determining} molecular embedding,~\cite{mol2vec,ma2018determining} molecular graph,~\cite{coley2017convolutional,ma2018determining} BIGSMILES.~\cite{lin2019bigsmiles}  Since our coarse-grained model only contains two types of beads, it is both pragmatic and appropriate to use one-hot encoding~\cite{Webb2020Target} to pre-process the polymer sequence information. As illustrated in Fig.~\ref{fig:ML_framework} (a), we encode the 1D 20-bead-length polymer chain's sequence information into a 20 dimensional vector, where type A beads are 1 and type B beads are 0. The resulting vector is the input for the SVR model, and is trained to reproduce the corresponding $\Delta F$ of each polymer chain that is obtained from the aforementioned biased MD simulations. The polymer sequence is treated as headless; rather than inserting this symmetry into the model, we augment the the dataset with this symmetric property, adding each sequence's backward representation with the same $\Delta F$ output unless that sequence is a palindrome.\footnote{To illustrate, if one polymer sequence is $[00110011001100110011]$ and the corresponding adhesive energy is $\Delta F_{\rm A}$, we can add its backward order sequence $[11001100110011001100]$ and $\Delta F_{\rm A}$ without running MD simulations. But if one polymer sequence is palindrome, like $[00000111111111100000]$ with $\Delta F_{\rm B}$, whose forward order and backward order are the same, only that sequence is used so as not to impart undue influence from that sequence to the SVR model.}

For each PS in Fig.~\ref{fig:model} (b-e), we collect $20,000$ polymer chains with unique sequences which have different compositions and different orders and the corresponding $\Delta F$. A separate SVR model is trained for each patterned surface. To train the SVR model, we use 5-fold cross-validation after shuffling the data, as shown in Fig.~\ref{fig:ML_framework} (b). We employ the coefficient of determination ($R^2$) and mean absolute error (MAE) to characterize the model's performance and optimize the SVR model's hyperparameters ($C$, $\gamma$ and $\epsilon$; see caption of Fig.~\ref{fig:ML_framework} (b)).

\begin{figure}[h]
	\begin{center}
	\includegraphics[width=\widefigurewidth]{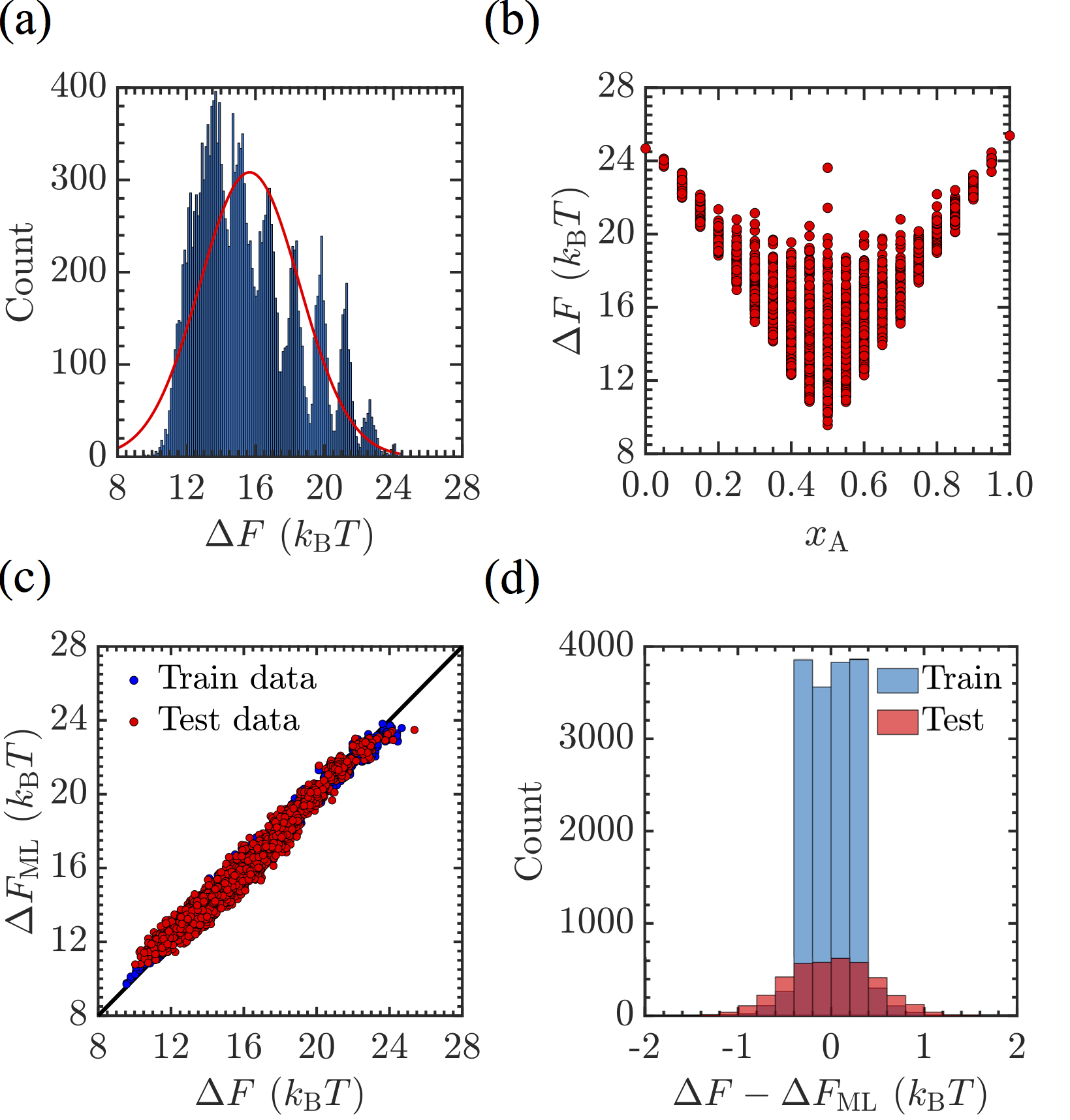}
	\end{center}
	\caption{Adhesive free energy data for the interaction between sequence-specified polymers and surface PS1. Panel (a) presents a histogram of $\Delta F$ illustrating the distribution of adhesive free energies of sequential polymer chains. A Gaussian fit with $\mu = 15.6634 \kT$ and $\sigma = 2.89355 \kT$ is shown for reference. Panel (b) depicts the distribution in $\Delta F$ with respect to the overall composition fraction $x_{\rm A}$ of the polymer. Panel (c) presents the training and predictive behavior of SVR models, with the predicted value  $\Delta F_{\rm ML}$  ($y$-axis) plotted against simulated value $\Delta F$ ($x$-axis) for the training data (blue) and test data (red). For the testing set, the $R^2$ score is $0.97277$ and  MAE is $0.3823 \kT$. Panel (d) shows a histogram of the deviation $\Delta F - \Delta F_{\rm ML}$ of the model from the true value, and demonstrates good predictive capability of our SVR model for this surface.
	}
	\label{fig:ps1}
\end{figure}

\section*{Results and Discussion}

We begin by exploring the performance of each surrogate model on the accumulated datasets for adhesive free energy. In Figures 4--7 we characterize the distribution of free energies within each dataset in panel (a), how diverse each free energy distribution is with respect to average composition (quantified by the fraction $x_{\rm A}$ of A-type monomers) in panel (b), and illustrate the errors in the training and test data in panels (c) and (d). Examining the data for PS1 (Fig.~\ref{fig:ps1}), we note that the distribution of adhesive free energies over the sequences space is quite broad, with the best binding occurring for nearly pure sequences ($x_A \approx 0$ or $x_A \approx 1$). Because the surface is symmetric in its placement of A and B-type beads, the distribution of binding energies is symmetric over the compositions $x_A$. As shown in Fig.~\ref{fig:ps1}(c, d), training data is clustered quite tightly around the SVR model, and all data within the test set is within $\approx 1 \kT$ of its actual adhesive value. For this case, the SVR is seen to give good accuracy and predictive capability.

\begin{figure}[h]
	\begin{center}
	\includegraphics[width=\widefigurewidth]{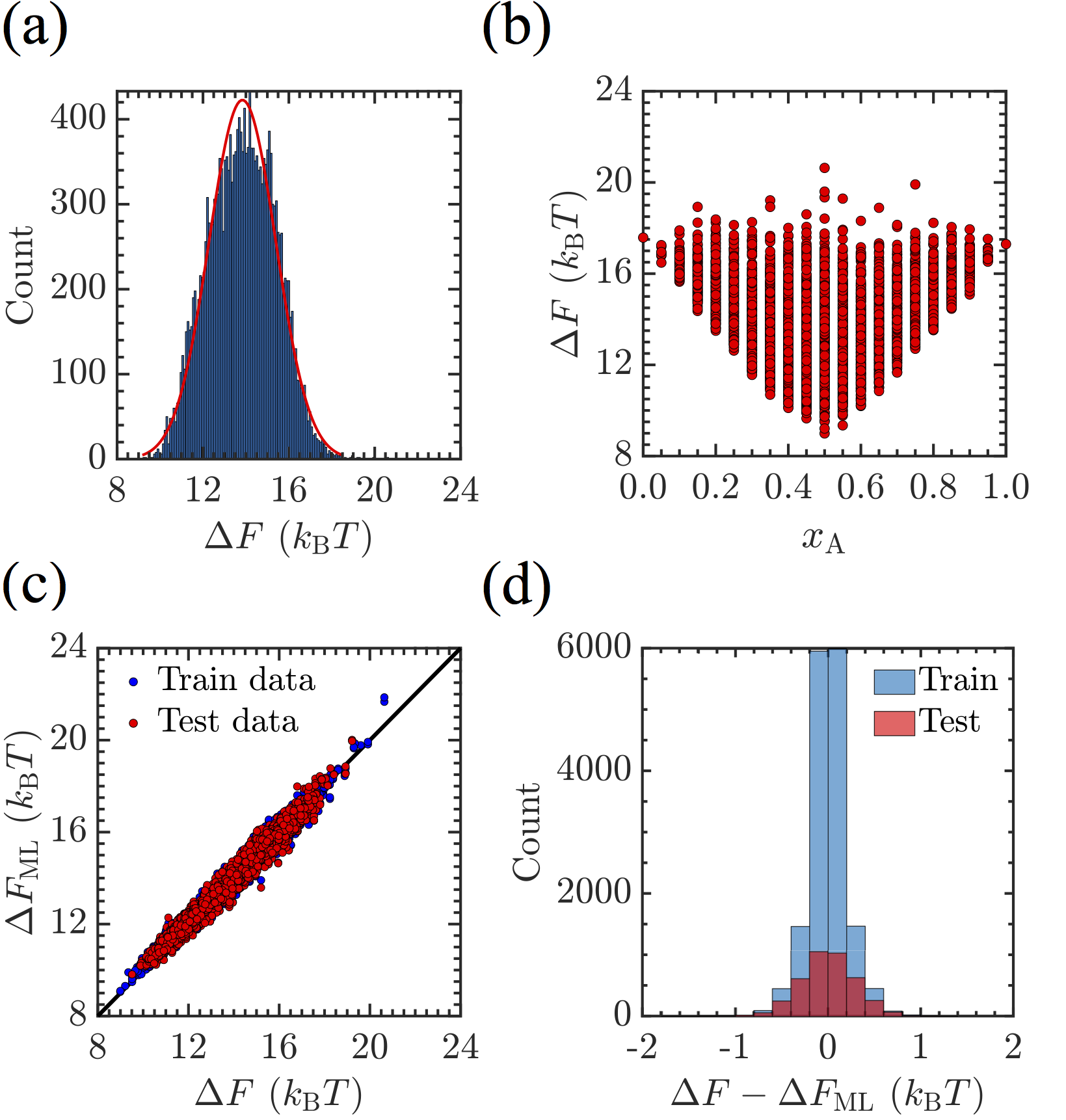}
	\end{center}
	\caption{Adhesive free energy data for the interaction between sequence-specified polymers and surface PS2. Panel (a) presents a histogram of $\Delta F$ illustrating the distribution of adhesive free energies of sequential polymer chains. A Gaussian fit with $\mu = 13.8378 \kT$ and $\sigma = 1.55364 \kT$ is shown for reference. Panel (b) depicts the distribution in $\Delta F$ with respect to the overall composition fraction $x_{\rm A}$ of the polymer. Panel (c) presents the training and predictive behavior of SVR models, with the predicted value  $\Delta F_{\rm ML}$  ($y$-axis) plotted against simulated value $\Delta F$ ($x$-axis) for the training data (blue) and test data (red). For the test set, the $R^2$ score is $0.96497$ and  MAE is $0.2300 \kT$. Panel (d) shows a histogram of the deviation $\Delta F - \Delta F_{\rm ML}$ of the model from the true value, and demonstrates good predictive capability of our SVR model for this surface.
	}
	\label{fig:ps2}
\end{figure}

Similar results are obtained for the surrogate model developed for surface PS2. This surface has more fine-grained structure, and this results in lower average binding energies, and a more unimodal distribution in interaction energies relative to PS1 [Fig~\ref{fig:ps2}(a)]. The distribution with respect to $x_{\rm A}$ is similarly more diffuse [Fig~\ref{fig:ps2}(b)]. Interestingly, the predictions for this surface are better than PS1 when the MAE is taken into account, and the test data is more tightly distributed indicating the SVR again has good predictive capability~[See Fig.~\ref{fig:ps2}(a,b) and caption].

\begin{figure}[h]
	\begin{center}
	\includegraphics[width=\widefigurewidth]{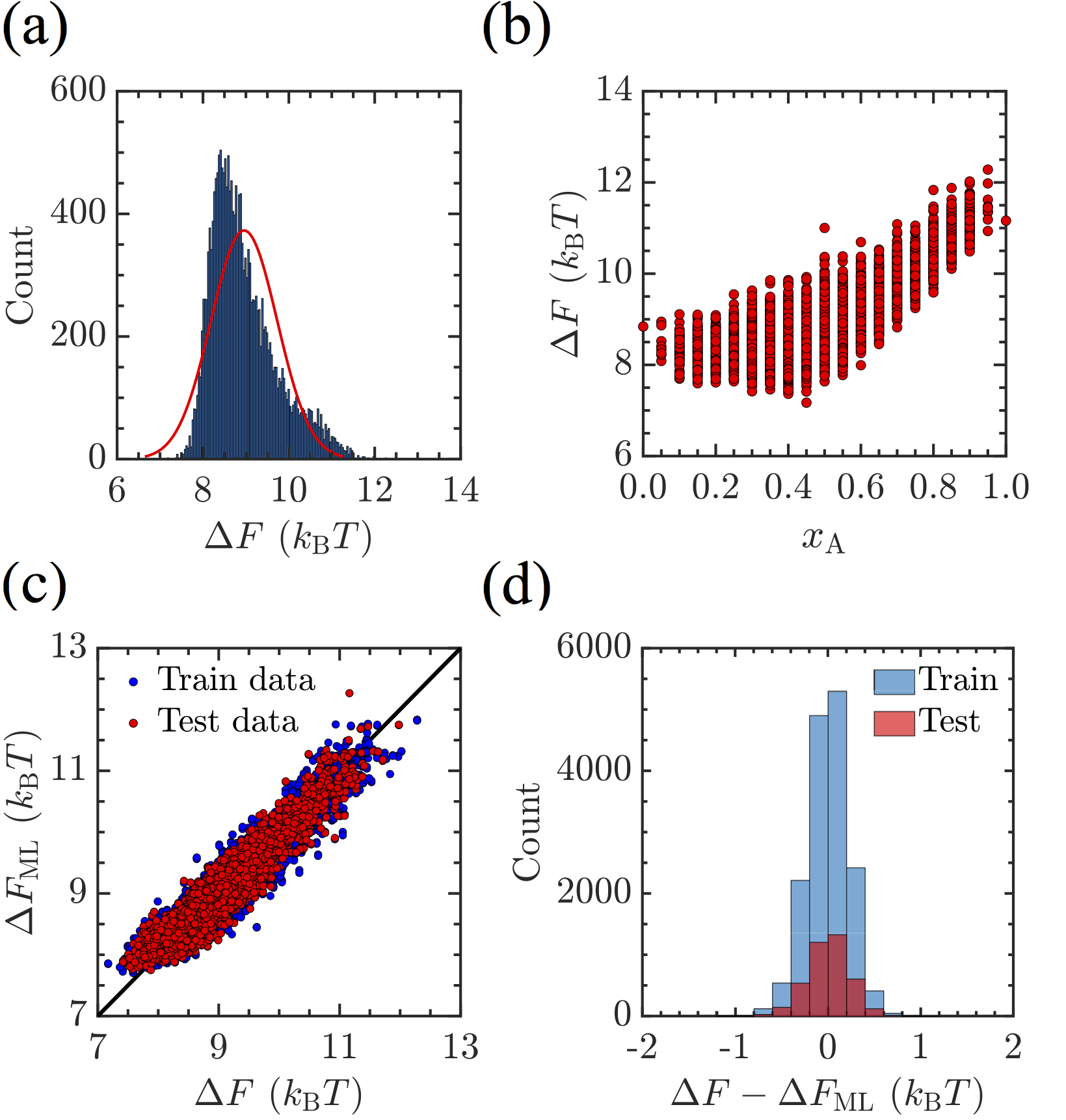}
	\end{center}
	\caption{Adhesive free energy data for the interaction between sequence-specified polymers and surface PS3. Panel (a) presents a histogram of $\Delta F$ illustrating the distribution of adhesive free energies of sequential polymer chains. A Gaussian fit with $\mu = 8.9575 \kT$ and $\sigma = 0.770151 \kT$ is shown for reference. Panel (b) depicts the distribution in $\Delta F$ with respect to the overall composition fraction $x_{\rm A}$ of the polymer. As this surface contains randomized elements, the distribution of adhesive energies is no longer symmetric. Panel (c) presents the training and predictive behavior of SVR models, with the predicted value  $\Delta F_{\rm ML}$  ($y$-axis) plotted against simulated value $\Delta F$ ($x$-axis) for the training data (blue) and test data (red). For the test set, the $R^2$ score is $0.90864$ and  MAE is $0.1796 \kT$. Panel (d) shows a histogram of the deviation $\Delta F - \Delta F_{\rm ML}$ of the model from the true value, and demonstrates good predictive capability of our SVR model for this surface, despite the broader distribution relative to the mean value of interaction energies with PS3 reltaive to PS2.
	}
	\label{fig:ps3}
\end{figure}

\begin{figure}[h]
	\begin{center}
	\includegraphics[width=\widefigurewidth]{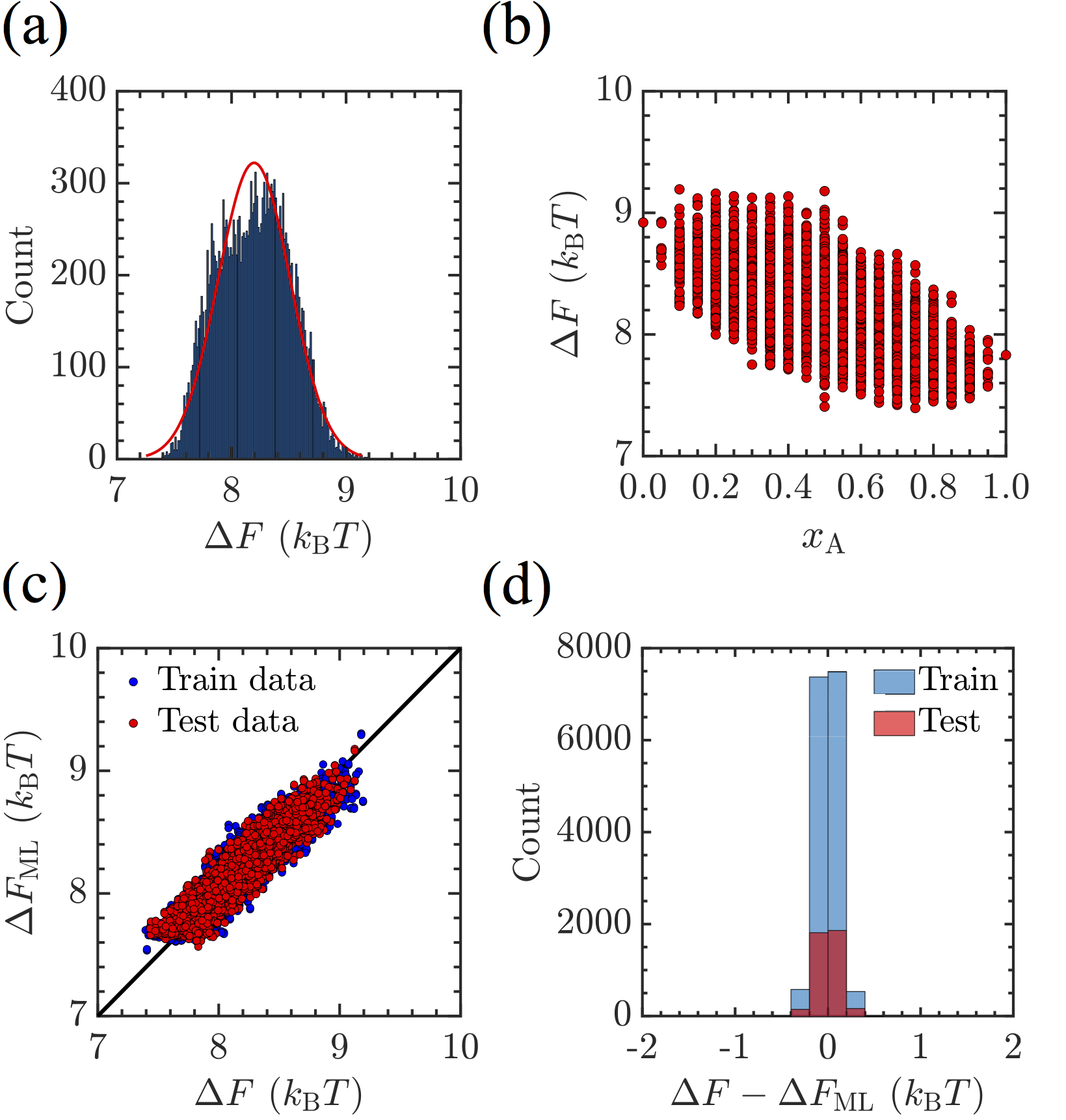}
	\end{center}
	\caption{Adhesive free energy data for the interaction between sequence-specified polymers and surface PS4. Panel (a) presents a histogram of $\Delta F$ illustrating the distribution of adhesive free energies of sequential polymer chains. A Gaussian fit with $\mu = 8.19659  \kT$ and $\sigma = 0.314477 \kT$ is shown for reference. Panel (b) depicts the distribution in $\Delta F$ with respect to the overall composition fraction $x_{\rm A}$ of the polymer. As this surface contains randomized elements, the distribution of adhesive energies is not symmetric. Panel (c) presents the training and predictive behavior of SVR models, with the predicted value  $\Delta F_{\rm ML}$  ($y$-axis) plotted against simulated value $\Delta F$ ($x$-axis) for the training data (blue) and test data (red). For the test set, the $R^2$ score is $0.86922$ and  MAE is $0.0903 \kT$. Panel (d) shows a histogram of the deviation $\Delta F - \Delta F_{\rm ML}$ of the model from the true value, and demonstrates exceptional predictive capability for this surface, with a narrower distribution of differences between test data and model prediction than the other surfaces studied.
	}
	\label{fig:ps4}
\end{figure}

PS3 contains an irregular, randomly generated surface structure where each bead's identity is equally probable to be A or B, so that the expected composition of the surface is $x_A=0.5$. Modeling results for this surface are shown in Fig.~\ref{fig:ps3}. The distribution is seen to have a narrower distribution and smaller mean in the energy distribution (Fig.~\ref{fig:ps3}(a)) than PS1 and PS2, resulting from the randomized features. Further, the randomization results in a skewed distribution of adhesive properties as a function of the composition (Fig.~\ref{fig:ps3}(b)). Despite this, the SVR model again performs extremely well, with $\mathcal{O}(\kT)$ accuracy in prediction, despite the relatively broad training set distribution (when compared to PS1 and PS2) [see Fig.~\ref{fig:ps3}(c,d)]. Similar effects are observed for surface PS4 (Fig.~\ref{fig:ps4}), which exhibits very high accuracy in predictions in absolute terms, and also when error relative to the mean interaction energy is considered.

From Fig.~\ref{fig:ps1}(a), Fig.~\ref{fig:ps2}(a), Fig.~\ref{fig:ps3}(a), and Fig.~\ref{fig:ps4}(a), we see that surface patterns can change both the distributions and the range of $\Delta F$. These differences from the training data would  be expected to affect the SVR models' prediction abilities. Though in some cases the $R^2$ is reduced, this is primarily to do with overall broadening of the distribution relative to the value of the mean of the distribution. We see that even for the poorest $R^2$ performance (PS4), the held-out test data still exhibits a high value of $0.86922$. Therefore, in general, we see that the SVR models have extremely good predictive capability.

\begin{figure}[h]
	\begin{center}
	\includegraphics[width=\widefigurewidth]{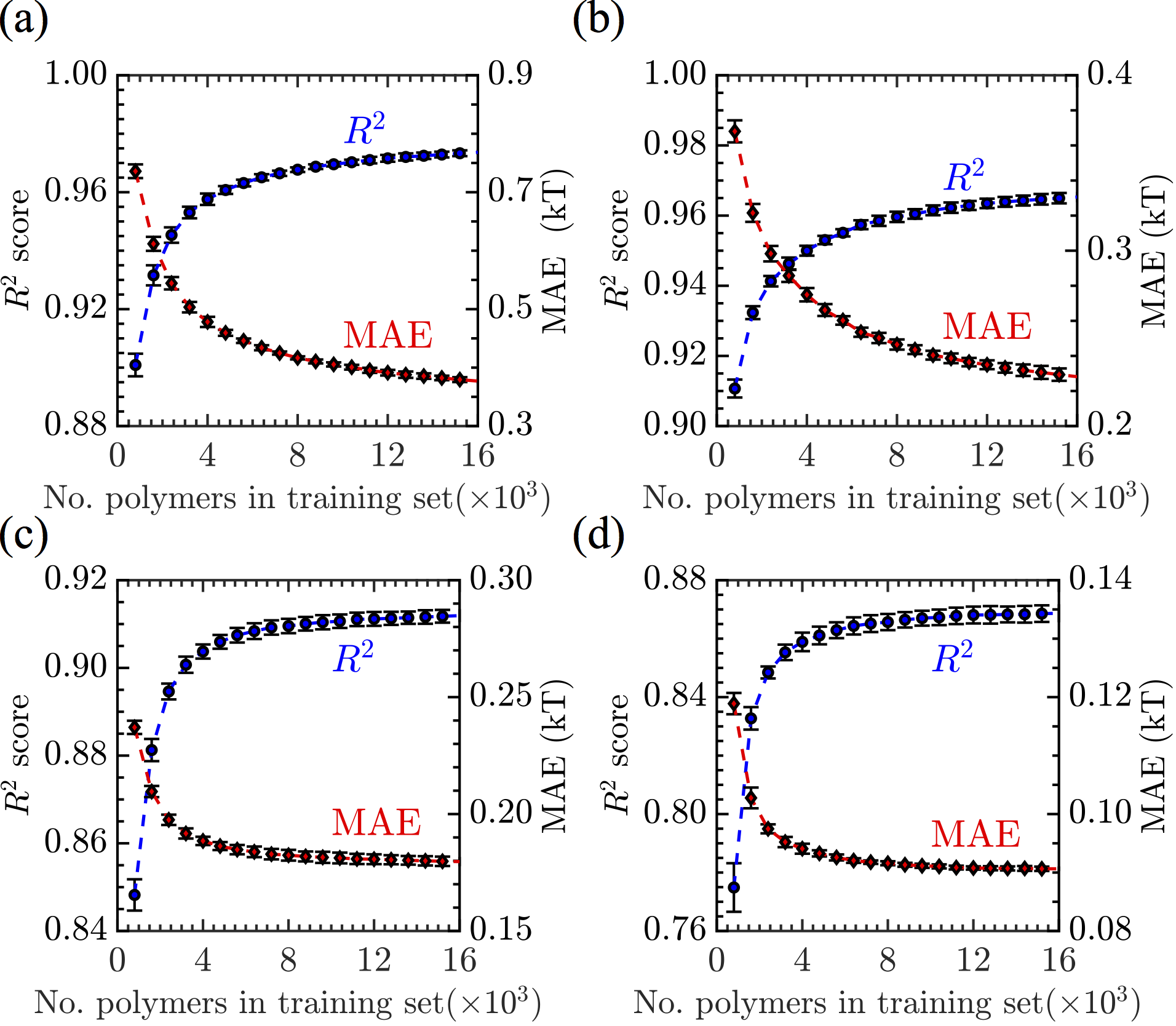}
	\end{center}
	\caption{$R^2$ score (blue circles, left axis) and MAE (red diamonds, right axis) associated with ML regression models when predicting $\Delta F$ as a function of the number of polymers in training set for PS1 (a), PS2 (b), PS3 (c), P4 (d). The error bars reflect the SD of the $R^2$ scores and the SD of the MAE from 5-fold cross-validation, in which each fold is used as a test set for SVR models trained using $5-100\%$ of the data from the remaining 4 folds. 
	}
	\label{fig:size}
\end{figure}

In the above investigations, we utilize $20,000$ data for each case. Typically, a larger dataset means better predictive accuracy for the model. However, it is difficult to obtain a large data sets in many cases due to the computational and experimental cost and potential complexity to create a database. Therefore, it is of interest to measure the quantity of data necessary to train an effective SVR model. Again, we use a 5-fold cross-validation strategy choose the training set size by randomly selecting $5-100\%$ of data from the remaining 4 folds. Fig.~\ref{fig:size} illustrates the performance of SVR models, as quantified by $R^2$ score (blue circles, left axis) and MAE (red diamonds, right axis), for predicting $\Delta F$ for the  held-out sequences ($4,000$) as a function of the training set size ($800-16,000$). The quality of the SVR models (as measured by $R^2$) for all the four surfaces improves monotonically as the training set size increases, with the sharpest increases coming prior to $8,000$ training data points; after this the $R^2$ levels off considerably. The mean absolute error continues to decrease after this, though the decrease is more pronounced for PS1 and PS2 than for PS3 and PS4. This implies that our model could have performed approximately as well with half of the data utilized here. The fact that only a few thousand data points seem to be required for models of this quality is promising, as it enables us to target the accuracy level and optimal size of the adhesion simulations used to populate our database. This fact also highlights a potential benefit of using ML, because the SVR machine learning models might be constructed from more limited data if the tolerances seen here are acceptable. Furthermore, the need for relatively few data points to get the essential character of polymer--interface interactions, along with the physical relations of each of the patterned surfaces to each other indicate transfer learning~\cite{ma2020transfer} could be a viable strategy to improve the model's accuracy in small datasets.

Finally, we show how our SVR models may be exploited to perform inverse-design  of polymer sequences with a desired $\Delta F$ by leveraging our SVR models trained on our previously generated data. Here, we apply a genetic algorithm to perform the optimization. We use an  initial population size of polymer sequences being $1,000$, a mating probability of $0.8$, and a mutation probability of $0.003$. The algorithm proceeds for $1,000$ steps. To search for the polymer sequences with the largest $\Delta F_{\rm ML}$, we set the fitness function so that larger $\Delta F_{\rm ML}$ has the larger fitness: 
\begin{equation}
  {\rm Fitness}_{i} = \Delta F^{i}_{\rm ML} - \min(\Delta F_{\rm ML}) + 0.001  \;.
\end{equation}
\noindent where $i$ is the index of the sequence, $\Delta F^{i}_{\rm ML}$ is the corresponding ML predicted adhesive free energy, ${\rm Fitness}_{i}$ is the corresponding fitness value. And $\min(\Delta F_{\rm ML})$ is the smallest ML predicted adhesive free energy among $1,000$ sequences in the current generation. The small numerical offset is applied so that the fitness function is always strictly greater than zero. Results are summarized in Table~\ref{tab:geneticmodel}. The target sequences for PS1, PS2, PS3 are already in the existing $20,000$ database, while the target sequence for PS4 is out of the existing $20,000$ database. 

\begin{center}
\begin{table}
\begin{tabular}{|c|c|c|c|}
\hline
Surface & Target Sequence & $\Delta F_{\rm ML}$ (\kT) & $\Delta F$ (\kT)\\
\hline
PS1 & $[0 0 0 0 0 0 0 0 0 0 0 0 0 0 0 0 0 0 0 0]$ & $23.742$ & $24.674$ \\
\hline
PS2 & $[0 0 0 0 0 0 0 0 0 0 1 1 1 1 1 1 1 1 1 1]$ & $21.680$ & $20.636$ \\
\hline
PS3 & $[1 1 1 1 1 1 1 1 1 1 1 1 1 1 1 1 1 1 1 1]$ & $12.273$ & $11.162$ \\
\hline
PS4 & $[0 0 0 0 0 0 0 0 0 0 0 0 1 1 1 1 1 1 1 1]^*$ & $9.359$ & $9.427$\\
\hline
\end{tabular}
\caption{\label{tab:geneticmodel} Target Sequence for Largest $\Delta F_{\rm ML}$ and corresponding $\Delta F$}
\end{table}
\end{center}

\begin{figure}[h]
	\begin{center}
	\includegraphics[width=\widefigurewidth]{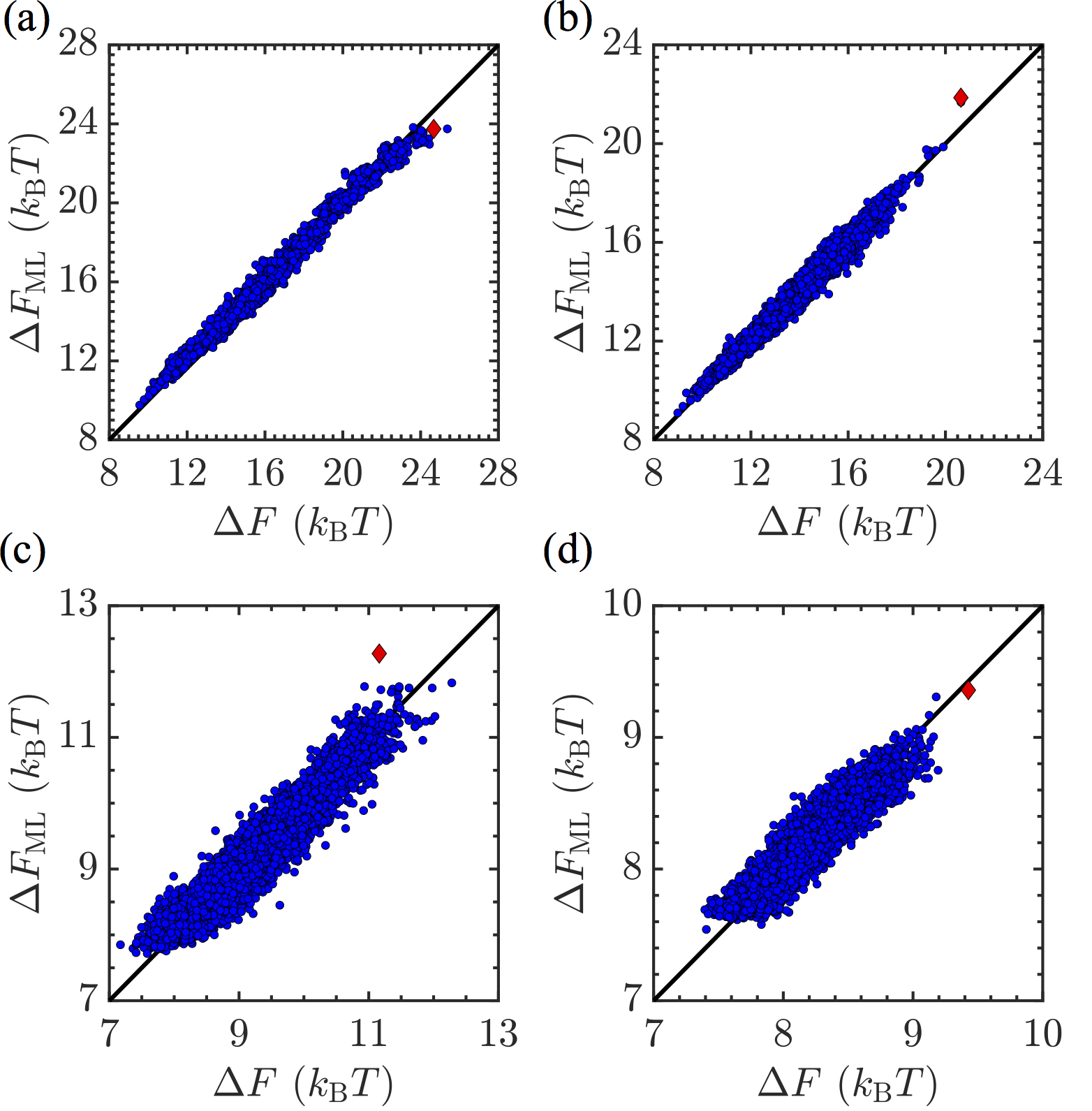}
	\end{center}
	\caption{The largest $\Delta F_{\rm ML}$ from application of a genetic algorithm (red diamond, see main text for description) vs the corresponding simulated $\Delta F$ is depicted against all sequences (blue circle) in the existing database for PS1 (a), PS2 (b), PS3 (c), PS4 (d).}
	\label{fig:GA}
\end{figure}
We find that the genetic algorithm gives good results, and use of the SVR model enables efficient searching of compositional space. Using the output string, we calculate the corresponding $\Delta F$ using MD simulations. Relative to the database, these are plotted using the red diamond in Fig.~\ref{fig:GA}. Though some small differences exist, these sequences are uniformly at the top end of the distribution in terms of maximal adhesion. One may improve these results, if desired, by first using genetic model to get the top $100$ sequences and then running MD simulations to get the true $\Delta F$ for these top $100$ sequences. From these outputs, the best $\Delta F$ may be chosen. Furthermore, if searching for other properties, like the sequence corresponding to smallest $\Delta F_{\rm ML}$ or one specific $\Delta F_{\rm ML}$ value, one needs only to modify the fitness function where the desired property have the largest fitness value. 

\section*{Conclusion}
We utilize a support vector regression ML model to predict the adhesion free energy $\Delta F$ of polymer-surface interaction with its sequence information as input. In our work, we test on four decorated surfaces with different patterns. The free energy ranges and energy distributions observed on the surfaces explored exhibit significant differences. The support vector regression ML model inexpensively and reliably predicts adhesive free energies of polymer-surface interactions from sequential information. Though the free energies for four different surfaces are very different, each model exhibits very good accuracy in prediction. We identify how similar accuracy may be obtained with slightly less data, and use the output of these models to design adhesive polymer sequences using a genetic algorithm, demonstrating good success on this inverse design problem.

Our work highlights the promising integration of coarse-grained simulation with data-driven machine learning methods for getting quantitative relationships between polymer sequences and adhesive free energies and  inverse-designing sequential polymers for pattern recognition. Our work thus represents a step forward from predicting the structural and functional properties of sequential polymer chains themselves to predicting their interactions with surfaces, enabling the design of polymer sequences for desired polymer--surface interactions.~\cite{gormley2021machine} While our molecular simulation model in this work is a toy coarse-grained model which only contains two types of backbone beads, the techniques of the data-driven machine learning workflow are readily generalized to more complex and realistic polymer chain models which can help mimic biological processes~\cite{xiu2020Inhibitors,chakraborty2001disordered,chakraborty2001polymer} and practical applications.\cite{kim2003epitaxial} Extensions of these studies to incorporate more specificity in the models and more general predictions of surface--polymer interactions represent targets for future work. As highlighted by the results here, there is reason to believe such refined strategies will be extremely successful in the creation of new adhesive materials.

\section*{Code Availability}
Example scripts and information necessary to run the examples contained in this article are posted at \url{https://github.com/shijiale0609/ML_PSI}.

\section*{Acknowledgement}
JS, MJQ, PA, and JKW acknowledge support through the Midwest Integrated Center for Computation of Materials (MICCoM), a DOE Computational Materials Science Center funded by the Department of Energy, Office of Science, Basic Energy Sciences, Materials Sciences and Engineering Division for the development and maintenance of the code SSAGES used to calculate free energies in this work. Additional support is acknowledged from startup funding provided by the University of Notre Dame, and computational resources at the Notre Dame Center for Research Computing (CRC). JS and JKW would like to acknowledge fruitful discussions with Ruimin Ma (University of Notre Dame) and Yamil Col\'{o}n (University of Notre Dame).

\bibliography{ref}

\end{document}